\documentstyle[12pt]{article}

\begin{document} 

\begin{titlepage}

\hrule 
\leftline{}
\leftline{Preprint
          \hfill   \hbox{\bf CHIBA-EP-105}}
\leftline{\hfill   \hbox{hep-th/9803133}}
\leftline{\hfill   \hbox{(revised)}}
\leftline{\hfill   \hbox{June 1998}}
\vskip 5pt
\hrule 
\vskip 1.0cm

\centerline{\large\bf 
Existence of Confinement Phase 
in Quantum Electrodynamics
} 
\vskip 0.5cm
\centerline{\large\bf  
}
\vskip 0.5cm
\centerline{\large\bf  
}

\vskip 1cm

\centerline{{\bf 
Kei-Ichi Kondo$^{1}{}^{\dagger}$
}}  
\vskip 4mm
\begin{description}
\item[]{\it  
$^1$ Department of Physics, Faculty of Science,
  Chiba University,  Chiba 263, Japan
  }
\item[]{$^\dagger$ 
  E-mail:  kondo@cuphd.nd.chiba-u.ac.jp 
  }
\end{description}
\vskip 0.5cm

\centerline{{\bf Abstract}} \vskip .5cm

We show that the four-dimensional U(1) gauge theory in the continuum
formulation has a confining phase (exhibiting area law of the
Wilson loop) in the strong coupling region above a critical coupling
$g_c$.  This result is obtained by taking into account topological
non-trivial sectors in U(1) gauge theory.  The derivation is based on
the reformulation of gauge theory as a deformation of a topological
quantum field theory and subsequent dimensional reduction of the
D-dimensional topological quantum field theory to the
(D-2)-dimensional nonlinear sigma model.  The topological
quantum field theory part of the four-dimensional U(1) gauge theory
is exactly equivalent to the two-dimensional O(2) nonlinear sigma
model.  The confining (resp. Coulomb) phase of U(1) gauge theory
corresponds to the high (resp. low) temperature phase of O(2)
nonlinear sigma model and the critical point $g_c$ is determined by
the Berezinskii-Kosterlitz-Thouless phase transition temperature. 
The quark (charge) confinement in the strong coupling phase is caused
by the vortex condensation.  Thus the continuum gauge theory has the
direct correspondence to the compact formulation of lattice gauge
theory.


\vskip 0.2cm
\hrule  

\vskip 0.5cm  

\begin{description}
\item[]{
$^*$ To be published in Phys. Rev. D.
 }  
\end{description}

\end{titlepage}

\pagenumbering{arabic}

\newpage
\section{Introduction}
\setcounter{equation}{0}
\par
\par
In this paper we study the phase structure of the {\it continuum}
Abelian U(1) gauge theory by including the effect
due to compactness of U(1) group. 
The reason of taking into compactness is as follows.
From the viewpoint of unified field
theory, the Abelian group should be embedded as a subgroup in larger
non-Abelian gauge group.  
In view of this,
the Abelian group should be compact. 
Another importance of compactness of
Abelian gauge group stems from a possibility of explaining
quantization of charge \cite{Polyakov87}. 
In non-compact QED there are no reasons for charge quantization.
\par
In this paper we show that the four-dimensional U(1) gauge theory has
a confinement phase in the strong coupling region $g>g_c$ due to the
compactness (periodicity) leading to the non-trivial topological
configuration. If we neglect the periodicity, we have a free U(1)
gauge theory which has only one phase, the Coulomb phase, as
expected. This work confirms the claim made by Polyakov
\cite{Polyakov75,Polyakov77}. However, the claim that the abelian
gauge theory has a confinement phase sounds strange from the
conventional wisdom based on the continuum abelian gauge theory. 
We clarify the meaning of this statement in what follows.
\par
More than twenty years ago, it has been pointed out by many authors
that the four-dimensional SU(2) non-Abelian gauge theory bears many
similarities with a two-dimensional O(3) nonlinear sigma model
(NLSM).  Both theories possess asymptotic freedom, multi-instanton
(and anti-instanton) solution, dynamical mass generation and scale
invariance (i.e., no intrinsic scale parameter), see \cite{KondoII}
for references.
\par
These similarities can be seen also in the lattice regularized
versions of these models, between spin models and lattice gauge
theories \cite{Wilson74}. Naively the scaling limit of the classical
O(3) Heisenberg model is the O(3) NLSM, whereas that of the SU(2)
lattice gauge theory is the SU(2) gauge theory.
One can take the scaling limit of a lattice theory at a second order
phase transition point.  Hence the scaling limit is taken by
approaching the critical point 
$T \rightarrow T_c$ (or $g \rightarrow g_c)$ as the lattice spacing
$a$ goes to zero,
$a \rightarrow 0$, in such a way that the physical quantities remain
finite.
\par
In two-dimensional classical O(3) Heisenberg model, the two-point
correlation function decays exponentially at any finite
temperature.  This corresponds to a claim in four-dimensional
lattice SU(2) gauge theory that confinement phase survives as long
as the coupling constant
$g$ is positive, even if $g \ll 1$. Both models have a phase
transition at
$T=0$ ($g=0)$, i.e., $T_c=0$ ($g_c=0)$ which is believed to be the
second order.
\par
In a previous paper \cite{KondoII}, it has been shown that these
similarities between two models are not merely  an accident;
actually we have proved the exact equivalence between
(D-2)-dimensional O(3) NLSM and the $D$-dimensional topological
quantum field theory (TQFT) obtained by removing the perturbative
deformation (topological trivial sector) from the
$D$-dimensional SU(2) non-Abelian gauge theory ($D\ge3$).   This
proof is based on an idea of the dimensional reduction of Parisi and
Sourlas \cite{PS79}. The case of
$D=4$ is the most interesting case of physical reality.

\par
What can we say in the Abelian case?
For this, recall a fact that the two-dimensional O(2)
NLSM or XY model undergoes a phase transition without the appearance
of a spontaneous magnetization.  This absence of order parameter in
two dimensions is consistent with the Coleman-Mermin-Wagner (CMW)
theorem \cite{CMW}. The low temperature phase ($T<T_c$) contains
massless spin waves.  On the other hand, the high temperature phase
($T>T_c$) is completely disordered.  For this phase transition, the
periodicity of the angular variable
$\varphi$ is quite essential.  The model has topological
singularities, called the vortices.  These vortices condense in
high temperature and disorder the correlation function \cite{BKT}. 
This phase transition is called the Berezinskii-Kosterlitz-Thouless
(BKT) transition
\cite{BKT}.  
The vortex part is equivalent to the neutral Coulomb gas and
sine-Gordon theory, see
\cite{Frohlich76,JKKN77,AGG80}.
Although the existence of BKT transition is rather subtle, the
existence of BKT transition was rigorously proved by Fr\"ohlich and
Spencer \cite{FS81}.
\par
In lattice formulation, it is well known that all of these properties
in two-dimensional abelian spin models have correspondences in the
Abelian gauge theory in four dimensions.  The vortices in two
dimensions are closely related to magnetic monopoles in four
dimensions, see  
\cite{BMK77,GJ77,Peskin78}.  The condensation of closed loops of
magnetic monopole currents leads to quark (charge) confinement in the
strongly coupled phase of Abelian gauge theory, since electric flux
cannot easily penetrate such a medium (which we call the dual
Meissner effect). It will be worth remarking that  the dual
superconductor vacuum of quantum chromodynamics (QCD) has been
derived recently without any {\it ad hoc} assumption
\cite{KondoI} from QCD in the continuum formulation.

\par
In lattice gauge theory, charge confinement in
the sense of area decay of the Wilson loop is derived in the strong
coupling region by using the strong coupling expansion
\cite{Wilson74,OS78}. 
Quite remarkably, the quark (charge) confinement in lattice gauge
theory occurs irrespective of the details of the gauge group, as
long as it is compact (discrete
\cite{EPS79,UWG80} or continuous), even for abelian gauge group. 
However, one expects that the U(1) lattice gauge theory in four
dimensions ($U(1)_4$) has a Coulomb phase in the weak coupling
region, which was proved rigorously by Guth
\cite{Guth80} and Fr\"ohlich and Spencer \cite{FS82}.
Therefore the U(1) lattice gauge theory undergoes a phase transition
at a finite non-zero coupling $g_c$.
In continuum gauge theory, such non-trivial phase
structure was suggested to occur due to topological
non-trivial configurations by Polyakov
\cite{Polyakov75,Polyakov87}.  
Actually he has shown the confinement phase in three-dimensional U(1)
gauge theory for arbitrary gauge coupling, in agreement with the
lattice analysis. In four dimensions, he claimed that the weak
coupling U(1) gauge theory does not confine.
Accordingly, it is expected that the U(1) gauge theory in four
dimensions have two phases, confinement and deconfinement (Coulomb)
phases, whereas only one phase i.e., confinement phase exists in
three dimensions. 
\par
In this paper, we show that the continuum four-dimensional U(1)
gauge theory has two phases, (strong coupling) confinement
phase and (weak coupling) Coulomb  phase, which have direct
correspondence with the high-temperature and low-temperature phases
in O(2) NLSM respectively. The phase transition point corresponds to
the BKT transition in XY model.  Therefore, the phase transition
point $g_c$ is determined by the BKT transition temperature $T_c$. 
This is one of the main results of this paper.
This result is obtained as a specific case of the previous paper
\cite{KondoII}.
\par
Therefore, in the strong coupling phase
($g>g_c$), the continuum U(1) gauge theory confines quarks and
the gauge field becomes massive, in agreement with the result of 
lattice gauge theory.  In the weak coupling phase, on the other
hand, quarks are liberated and gauge field remains massless. In the
weak coupling phase ($g<g_c$), the beta function of the
renormalization group  \cite{WK74} is identically zero and
$0<g<g_c$ is the line of fixed points, if we neglect the perturbative
deformations.  For a review of lattice gauge theory, see e.g.
\cite{Kogut79,DZ83,Seiler82}. 
\par
The plan of this paper is as follows.  In section 2, we give the
reformulation of Abelian gauge theory as a deformation of a
topological quantum field theory.  
In section 3, using the reformulation of section 2, we evaluate the
Wilson loop expectation value in four-dimensional Abelian gauge
theory.  In the final section we discuss the
renormalization group properties and extension of our scheme to other
dimensional cases.
More details about the interplay between the Abelian and non-Abelian
cases are given in a forthcoming paper.

\section{Abelian gauge theory as a deformation of TQFT and
dimensional reduction}

\par
Now we reformulate quantum electrodynamics (QED) as a deformation of
a topological quantum field theory.
It is obtained as a special case of non-Abelian gauge theory given
in the previous paper \cite{KondoII}.

\subsection{Decomposition into perturbative and topological
non-trivial sectors}
\par
 QED on the $D$-dimensional space-time is
defined by the action,
\begin{eqnarray}
 S_{QED}^{tot}  &=& \int d^Dx ({\cal L}_{QED}[a_\mu,\psi]
 + {\cal L}_{GF}),
 \\
 {\cal L}_{QED}[a_\mu,\psi] &:=& -{1 \over 4} 
  f_{\mu\nu} f_{\mu\nu} 
 + \bar \psi (i \gamma^\mu D_\mu[a] - m) \psi ,
\end{eqnarray}
where 
\begin{eqnarray}
 f_{\mu\nu}(x) 
&:=&    \partial_\mu a_\nu(x) 
 -   \partial_\nu a_\mu(x) ,
  \\
  D_\mu[a] &:=& \partial_\mu - i g a_\mu .
\end{eqnarray}
\par
The gauge transformation of the U(1) gauge field $a_\mu(x)$ and the
fermion field $\psi$ is defined by
\begin{eqnarray}
 a_\mu(x) &\rightarrow&  
 a_\mu^U(x) := 
  a_\mu(x)  + {i \over g} U(x) \partial_\mu U^\dagger(x) , \quad
   U(x)  \in U(1),
   \\
  \psi(x) &\rightarrow& \psi^U(x) := U(x) \psi(x) .
\label{GT0a}
\end{eqnarray}
The gauge fixing term ${\cal L}_{GF}$ is given by
\begin{eqnarray}
 {\cal L}_{GF} := - i \delta_B G_{gf}[a_\mu, C, \bar C, \phi] ,
 \label{GFa}
\end{eqnarray}
using  the nilpotent Becchi-Rouet-Stora-Tyupin (BRST) transformation
$\delta_B$,
\begin{eqnarray}
   \delta_B a_\mu(x)  &=&   \partial_\mu C(x)  ,
    \nonumber\\
   \delta_B C(x)  &=& 0 ,
    \nonumber\\
   \delta_B \bar C(x)  &=&   i \phi(x)  ,
    \nonumber\\
   \delta_B \phi(x)  &=&  0 ,
    \nonumber\\
   \delta_B \psi(x)  &=&   ig C(x) \psi(x) , \quad
   \delta_B \bar \psi(x)  =   - ig C(x) \bar \psi(x) ,
    \label{BRST0a}
\end{eqnarray}
where $\phi$ is the Lagrange multiplier field.
\par
The partition function of QED with the source term
\begin{eqnarray}
   S_J[a_\mu, C, \bar C, \phi, \psi, \bar \psi]
   := \int d^Dx ( J^\mu a_\mu + J_c C 
   + J_{\bar c} \bar C + J_\phi \phi +
\bar \eta \psi + \eta \bar \psi ) 
\end{eqnarray}
is given by
\begin{eqnarray}
 Z_{QED}[J] := \int [da_\mu][dC][d\bar C]
 [d\phi][d\psi][d\bar \psi] \exp \left\{ i S_{QED}^{tot} + i S_J
\right\} .
\end{eqnarray}

\par
To reformulate QED as a deformation of a topological quantum field
theory according to \cite{KondoII}, we first regard the U(1) gauge
field
$a_\mu$ and the fermion field $\psi$ as the gauge transformation of
the U(1) gauge fields $v_\mu$ and  $\Psi$,
\begin{eqnarray}
 a_\mu(x) &:=&   v_\mu(x)  + \omega_\mu(x),
   \quad \omega_\mu(x) := {i \over g}
U(x) \partial_\mu U^\dagger(x) . 
\\
  \psi(x)  &:=&  U(x) \Psi(x), 
\label{GT3a}
\end{eqnarray}
Here 
\footnote{
 The decomposition of $a_\mu$,
 $a_\mu =  v_\mu  + \omega_\mu$, 
corresponds to the superposition of two independent
configuration, $\varphi=\varphi_{SW}+\varphi_{V}$ (spin waves and
vortex parts)  in XY model.
}
$v_\mu$ and $\Psi$ are identified with the field variables in the
perturbative (topological trivial) sector ($Q=0$), whereas
$\omega_\mu$ belongs to the topological nontrivial sector
($Q\not=0$).
\par
Furthermore we introduce new ghost field $\gamma$, anti-ghost
field $\bar \gamma$ and the Lagrange multiplier field $\beta$ in the
perturbative sector.  They are subject to a new BRST
transformation 
$\tilde \delta_B$,
\begin{eqnarray}
   \tilde \delta_B v_\mu(x)  &=&   \partial_\mu \gamma(x)   ,
    \nonumber\\
   \tilde \delta_B \gamma(x)  &=&  0 ,
    \nonumber\\
   \tilde \delta_B \bar \gamma(x)  &=&   i \beta(x)  ,
    \nonumber\\
   \tilde \delta_B \beta(x)  &=&  0 ,
    \nonumber\\
   \tilde \delta_B \Psi(x)  &=&  ig \gamma(x) \Psi(x)  , \quad
   \tilde \delta_B \bar \Psi(x)  =   - ig \gamma(x) \bar \Psi(x) ,
    \label{BRST1a}
\end{eqnarray}
Then the partition function of QED is rewritten as
\begin{eqnarray}
 Z_{QED}[J] &=& \int [dU][dC][d\bar C]
 [d\phi]
 \int [dv_\mu][d\gamma][d\bar \gamma][d\beta]
 [d\Psi][d\bar \Psi] 
 \nonumber\\&&
\times \exp \Big\{ i \int d^Dx \Big(
 -i \delta_B G_{gf}[\omega_\mu +   v_\mu , C, 
 \bar C, \phi] \Big) 
 \nonumber\\&& \quad \quad \quad \quad
 + i \int d^Dx \Big( {\cal L}_{QED}[v,\Psi]  
-i \tilde \delta_B \tilde G_{gf}[v_\mu, \gamma, \bar \gamma,
\beta] \Big) 
 \nonumber\\&& \quad \quad \quad \quad
 + i S_J[\omega_\mu +  v_\mu, C, \bar C, \phi,  
  U\Psi, \bar \Psi U^\dagger ]
\Big\} ,
\label{formulaa}
\end{eqnarray}
where $\tilde G_{gf}$ is a gauge fixing functional for the
perturbative (topological trivial) sector.

\subsection{Gauge fixing}
\par
The Lorentz gauge is given by
\begin{eqnarray}
   F[a] := \partial_\mu a^\mu = 0 .
\end{eqnarray}
The most familiar choice of $G_{gf}$ 
\begin{eqnarray}
  G_{gf} = \bar C(\partial_\mu a^\mu + {\alpha \over 2}\phi) 
\end{eqnarray}
 yields the familiar form of the gauge fixing term,
\begin{eqnarray}
 {\cal L}_{GF} &:=& - i \delta_B G_{gf}[a_\mu, C, \bar C, \phi] 
 =   \phi \partial_\mu a^\mu + i \bar C
\partial^\mu \partial_\mu C + {\alpha \over 2} \phi^2  .
\label{gf0}
\end{eqnarray}
\par
In this paper we propose to use a choice,
\begin{eqnarray}
G_{gf}^{U(1)}
= - \bar \delta_B \left( {1 \over 2} a_\mu^2
+ i C \bar C \right) ,
\label{gf1}
\end{eqnarray}
where $\bar \delta_B$ is the anti-BRST transformation \cite{KondoII},
\begin{eqnarray}
   \bar \delta_B a_\mu(x)  &=&  \partial_\mu \bar C(x)  ,
    \nonumber\\
   \bar \delta_B  C(x)  &=&   i \bar \phi(x)  ,
    \nonumber\\
   \bar \delta_B \bar C(x)  &=& 0 ,
    \nonumber\\
   \bar \delta_B \bar \phi(x)  &=&  0 ,
    \nonumber\\
  \bar \delta_B \psi(x)  &=&    \bar C(x) \psi(x) ,
    \nonumber\\
\phi(x) + \bar \phi(x) &=& 0 ,
    \label{aBRSTa}
\end{eqnarray}
where $\bar \phi$ is defined in the last equation.
\par
Apart from a total derivative term, this choice yields
\begin{eqnarray}
{\cal L}_{GF} 
= i \delta_B \bar \delta_B \left( {1 \over 2} a_\mu^2
+ i C  \bar C  \right)
= - i \delta_B [\bar C(\partial_\mu a^\mu - \phi)] .
\end{eqnarray}
Therefore, the choice (\ref{gf1}) corresponds in (\ref{gf0}) to the
choice of the gauge-fixing parameter,
\begin{eqnarray}
 \alpha = -2 ,
\end{eqnarray}
which has appeared also
in the non-Abelian case
\cite{KondoII}.
The above choice for $G_{gf}^{U(1)}$ yields the decomposition,
\begin{eqnarray}
{\cal L}_{GF} 
&=& -i \delta_B G_{gf}^{U(1)}[\omega_\mu +  v_\mu , C,  \bar C, \phi]
 \\ 
&=&  i \delta_B \bar \delta_B \left( {1 \over 2} 
(\omega_\mu +  v_\mu)^2
+ i C  \bar C  \right)
 \nonumber\\ 
 &=& {\cal L}_{TQFT}
 +  i v_\mu  \delta_B \bar \delta_B \omega_\mu ,
 \label{conda}
\end{eqnarray}
where we have  defined
\begin{eqnarray}
 {\cal L}_{TQFT} &:=&  i \delta_B \bar \delta_B \left( {1 \over 2}
\omega_\mu^2 + i C  \bar C  \right) .
 \label{condaa}
\end{eqnarray}
Here we have used that the action of $\delta_B$ is trivial in the
perturbative sector, 
\begin{eqnarray}
 \delta_B v_\mu= 0 = \bar \delta_B v_\mu  ,
\end{eqnarray}
while
\begin{eqnarray}
 \delta_B \omega_\mu = \partial_\mu C,
 \quad \bar \delta_B \omega_\mu  = \partial_\mu \bar C .
\end{eqnarray}

\subsection{Deformation of topological quantum field theory}
\par
Finally, the partition function of QED is cast into the form,
\begin{eqnarray}
 Z_{QED}[J] &:=& \int [dU][dC][d\bar C]
 [d\phi]
\exp \Big\{ i S_{TQFT}[\omega_\mu, C, \bar C, \phi] 
 \nonumber\\&&   
  +  i \int d^Dx  [J^\mu \omega_\mu + J_c C 
   + J_{\bar c} \bar C + J_\phi \phi]  
+ i W[U; J^\mu, \bar \eta, \eta]  \Big\} ,
\end{eqnarray}
where $W[U; J^\mu, \bar \eta, \eta]$ is the generating functional of
QED  in the perturbative sector (pQED) given by
\begin{eqnarray}
 e^{i W[U; J^\mu, \bar \eta, \eta]}
&:=& 
   \int [dv_\mu][d\gamma][d\bar \gamma][d\beta]
 [d\Psi][d\bar \Psi] 
 \exp \Big\{ 
 i S_{pQED}[v,\Psi,\gamma, \bar \gamma, \beta]
\nonumber\\&&  
 +  i \int d^Dx \Big[  v_\mu {\cal J}_\mu 
 +  \bar \eta U \Psi + \eta \bar \Psi U^\dagger \Big] 
\Big\} ,
\\
 S_{pQED}[v,\Psi,\gamma, \bar \gamma, \beta] &:=& \int d^Dx
\Big[ {\cal L}_{QED}[v,\Psi]   -i \tilde \delta_B \tilde
G_{gf}(v_\mu, \gamma, \bar \gamma, \beta) \Big]  ,
\\
{\cal J}_\mu  &:=&   J_\mu   + i\delta_B \bar \delta_B \omega_\mu .
\end{eqnarray}
The correlation functions of the original (fundamental) field
$a_\mu, \psi, \bar \psi$ are
obtained by differentiating $Z_{QED}[J]$ with respect to the
corresponding source 
$J_\mu, \bar \eta, \eta$.  
\par
All the field configurations are classified
according to the integer-valued topological charge $Q$ which is
specified later. The above reformulation of gauge theory is the
decomposition of the original theory into the topological trivial
sector with
$Q=0$ and topological non-trivial sector with $Q\not=0$.
This corresponds to the decomposition of XY model into spin wave
part ($Q=0$) and the vortex part ($Q\not=0$) where $Q$ is given by
the winding number of the vortex solution.  However, the XY model
is not a gauge theory and does not have any local gauge invariance.
\par
The integration over the fields 
$(U, C, \bar C,\phi)$ in TQFT should be treated
non-perturbatively by taking into account the topological
non-trivial configurations.  
The deformation  $W[U; J^\mu, \bar \eta, \eta]$ from the TQFT should
be calculated according to the ordinary perturbation theory in the
coupling constant $g$. 
The perturbative expansion around the
TQFT means the integration over the new fields 
$(v_\mu, \gamma, \bar \gamma, \beta)$ based on the
perturbative expansion in powers of the coupling constant $g$.

\subsection{Dimensional reduction to O(2) NLSM}

Following the argument given in \cite{KondoII} based on the
Parisi-Sourlas dimensional reduction, it turns out that the
D-dimensional TQFT (as the topological non-trivial sector of
D-dimensional U(1) abelian  gauge theory) with an action
\begin{eqnarray}
 S_{TQFT}[\omega_\mu, C, \bar C, \phi] 
 = \int d^{D}x \
 i \delta_B \bar \delta_B \left(
 {1 \over 2}\omega_\mu(x) \omega_\mu(x)
+ i C(x) \bar C(x) \right)
\label{TQFT}
\end{eqnarray}
is equivalent to the
(D-2)-dimensional O(2) NLSM with an action,
\begin{eqnarray}
 S_{O(2)NLSM}[U] &:=& 2\pi  \int d^{D-2}z \ 
 {1 \over 2} \omega_\mu(z) \omega_\mu(z) , \quad 
\omega_\mu(z) := {i \over g} U(z) \partial_\mu U^\dagger(z) ,
\\
&=&  \int d^{D-2}z \ {\pi \over g^2}
  \partial_\mu U(z) \partial_\mu U^\dagger(z) .
\end{eqnarray}
Dimensional reduction is due to a fact that the
action (\ref{TQFT}) has a hidden supersymmetry and can be rewritten
in the $OSp(D/2)$ symmetric form in the superspace formulation, see 
\cite{KondoII}.

\section{Quark confinement in Abelian gauge theory}
\setcounter{equation}{0}

Now we calculate the Wilson loop expectation in the U(1) gauge
theory based on the reformulation given in the previous section.
In what follows we move to the Euclidean formulation.

\subsection{Dimensional reduction of Wilson loop}

We define the Wilson loop operator for the closed loop $C$ by
\begin{eqnarray}
 W_C[a] = \exp \left( i q \oint_C a_\mu(x) dx^\mu \right) ,
\end{eqnarray}
where $q$ is a test charge.
In the abelian gauge theory, the Wilson loop factorizes,
\begin{eqnarray}
 W_C[a] 
 = \exp \left( i q \oint_C \omega_\mu(x) dx^\mu \right)
 \exp \left( i q \oint_C v_\mu(x) dx^\mu \right) 
 =: W_C[\omega] W_C[v] .
\end{eqnarray}
\par
For $Q=0$ sector, we choose the gauge-fixing function,
 \begin{eqnarray}
 \tilde G_{gf} = \bar \gamma(\partial_\mu v^\mu + {\xi \over 2}
\beta ),
\end{eqnarray} 
with a gauge-fixing parameter $\xi$.
 For the U(1) gauge theory with an action  (omitting
matter fields), 
\begin{eqnarray}
 S_{pU(1)}[v,\gamma, \bar \gamma, \beta]
 := {1 \over 4g^2} (\partial_\mu v_\nu - \partial_\mu v_\nu)^2 
 - i \tilde \delta_B \tilde G_{gf} ,
\end{eqnarray} 
the perturbative part is given by 
\begin{eqnarray}
 e^{iW[\omega,J,0,0]} &=& \int [dv_\mu][d\gamma][d\bar
\gamma][d\beta]
 \exp \Big\{ 
 - S_{pU(1)}[v,\gamma, \bar \gamma, \beta]
 +  i \int d^Dx   v_\mu {\cal J}_\mu     \Big\} .
\end{eqnarray} 
Integrating out the fields $\gamma, \bar \gamma,  \beta$ yields
\begin{eqnarray}
 e^{iW[\omega,J,0,0]} =
 \int [dv_\mu]  \exp \Biggr\{ - S[v] - \int d^Dx  i   v_\mu(x) 
 [J_\mu(x)   + i\delta_B \bar \delta_B \omega_\mu(x)] \Biggr\},
\end{eqnarray} 
where
\begin{eqnarray}
 S[v] := \int d^Dx \left[
 {1 \over 4g^2} (\partial_\mu v_\nu - \partial_\mu v_\nu)^2
 + {1 \over 2\xi}(\partial_\mu v^\mu)^2  \right] .
\end{eqnarray} 
\par
For the calculation of the Wilson loop expectation, $J^\mu$ is
taken to be the current along the closed loop $C$ such that
\begin{eqnarray}
 \int d^D x v_\mu(x) J^\mu(x) = q \oint v_\mu(x) dx^\mu .
\end{eqnarray}
It is easy to see that
\begin{eqnarray}
 e^{iW[\omega,J,0,0]} =  \langle W_C[v]
 e^{(v_\mu, \delta_B \bar \delta_B \omega_\mu)} \rangle_{pU(1)}
   \int [dv_\mu]  e^{ - S[v] +      
   (v_\mu,  \delta_B \bar \delta_B \omega_\mu) } ,
\end{eqnarray} 
where
\begin{eqnarray}
 (v_\mu, \delta_B \bar \delta_B \omega_\mu)
:= \int d^Dx v_\mu(x) \delta_B \bar \delta_B  \omega_\mu(x) .
\end{eqnarray} 
The Wilson loop expectation is rewritten as
\begin{eqnarray}
  \langle W_C[a]  \rangle_{U(1)}
 = {\langle W_C[\omega] e^{iW[\omega,J,0,0]}  \rangle_{TQFT} \over 
 \langle  e^{iW[\omega,0,0,0]}  \rangle_{TQFT}} ,
  \label{wle1}
\end{eqnarray} 
where
\begin{eqnarray}
 e^{iW[\omega,0,0,0]} &=&  \langle  
 e^{(v_\mu, \delta_B \bar \delta_B \omega_\mu)} \rangle_{pU(1)}
   \int [dv_\mu]  e^{ - S[v] +      
   (v_\mu,  \delta_B \bar \delta_B \omega_\mu) } .
\end{eqnarray} 
Expanding the exponential 
\begin{eqnarray}
e^{(v_\mu, \delta_B \bar \delta_B \omega_\mu)}
= e^{ \int d^Dx \delta_B \bar \delta_B (v_\mu \omega_\mu)(x)}
  = e^{ \int d^Dx \{Q_B, \bar \delta_B (v_\mu
\omega_\mu)(x)\} }
\end{eqnarray} 
into power series and using a fact that the vacuum of TQFT obeys
\begin{eqnarray}
Q_B |0\rangle_{TQFT} = 0 , 
\end{eqnarray} 
we find that this term does not contribute to the expectation value
(\ref{wle1}). Therefore, the Wilson loop expectation is completely
separated into the topological part ($Q\not=0$) and the perturbative
part ($Q=0$),
\begin{eqnarray}
  \langle W_C[a]  \rangle_{U(1)}
 = \langle W_C[\omega] 
 \langle W_C[v]  \rangle_{pU(1)} \rangle_{TQFT}
 =  \langle W_C[\omega] \rangle_{TQFT}
 \langle W_C[v]  \rangle_{pU(1)} .
\end{eqnarray} 
This corresponds to the result eq.(10) of Polyakov \cite{Polyakov75}.
This property does not hold in the non-Abelian case, which makes the
systematic calculation rather difficult.
\par
In order to use the dimensional reduction for calculating the
topological part, we choose the loop
$C$ so that $C$ is contained in the $D-2$ dimensional
space.   For
$D=4$, the loop $C$ must be planar.
Then the dimensional reduction of the topological part leads to the
equivalence of the Wilson loop expectation value between the
D-dimensional U(1) TQFT and (D-2)-dimensional O(2) NLSM,
\begin{eqnarray}
  \langle W_C[\omega(z)] \rangle_{TQFT_D}
  = \langle W_C[\omega(z)] \rangle_{O(2) NLSM_{D-2}} ,
  \quad z \in R^{D-2} .
\end{eqnarray}
if the Wilson loop has its support on the (D-2)-dimensional space on
which the NLSM is defined.
Hence, the calculation of the Wilson loop
in the four-dimensional U(1) gauge theory is reduced to  those in the
two-dimensional O(2) NLSM  and the four-dimensional perturbative
U(1) gauge theory, 
\begin{eqnarray}
  \langle W_C[a]  \rangle_{U(1)_D}
 =  \langle W_C[\omega] \rangle_{O(2) NLSM_{D-2}}
 \langle W_C[v]  \rangle_{pU(1)_D} .
 \label{iden}
\end{eqnarray} 
\par
For large rectangular loop with sides $R, T$, the static potential is
obtained by 
\begin{eqnarray}
  V(R) = \lim_{T\rightarrow \infty} {-1 \over T} \ln \langle W_C[a] 
\rangle .
\end{eqnarray} 
In four dimensions ($D=4$), it is well known \cite{Kogut79} that for
large Wilson loop 
$\langle W_C[v]  \rangle_{pU(1)_4}$ gives the Coulomb potential, 
\begin{eqnarray}
   V(R) = - {g^2 \over 4\pi} {1 \over R} + {\rm constant} .
   \label{Cp}
\end{eqnarray} 
For a derivation, see e.g. Appendix of \cite{KondoIV}.

\par
In the following, we show that 
$\langle W_C[\omega] \rangle_{O(2) NLSM_{2}}$ exhibits area law for
strong coupling $g>g_c$ with a finite and non-zero value of a
critical point $g_c$.  This confinement-deconfinement transition
corresponds exactly to the BKT transition.

\subsection{O(2) NLSM and Wilson loop}

Defining the angle variable $\varphi(z)$ for $U(z) \in U(1)$,
\begin{eqnarray}
U(z) = e^{i\varphi(z)}  , 
\end{eqnarray}
we obtain
\begin{eqnarray}
  \omega_\mu(z) = {i \over g} U(z) \partial_\mu U^\dagger(z)
 = {1 \over g} \partial_\mu \varphi(z) . 
\end{eqnarray}
Then the action of O(2) NLSM reads
\begin{eqnarray}
 S_{O(2)}[U]  =  \int d^{D-2}z \pi \omega_\mu(z) \omega_\mu(z)
 = {\beta \over 2} \int d^{D-2}z \partial_\mu \varphi(z)
 \partial_\mu \varphi(z),
 \quad \beta := {2\pi \over g^2} .
 \label{fs}
\end{eqnarray}
The partition function is defined by
\begin{eqnarray}
 Z_{O(2)}[U]  
 = \int [dU] \exp (- S_{O(2)}[U]), \quad
 [dU] = \prod_{x} {d\varphi(x) \over 2\pi} ,
\end{eqnarray}
using the Haar measure $dU$ on U(1).
For the notation of the two-dimensional vector,
\begin{eqnarray}
 {\bf S}(x) = (\cos \varphi(x), \sin \varphi(x)) ,
\end{eqnarray}
the action reads
\begin{eqnarray}
 S_{O(2)}[U]  
 = {\beta \over 2} \int d^{D-2}z \
 \partial_\mu {\bf S}(x) \cdot \partial_\mu {\bf S}(x) .
\end{eqnarray}
Hence the O(2) NLSM is regarded as a continuum version of classical
planar spin model.
In the following, we identify $\beta$ with the inverse temperature
$1/T$.  Hence the high (resp. low) temperature of the spin
model corresponds to strong (resp. weak) coupling of the gauge
theory. 
\par
It should be remarked that O(2) NLSM with the
action (\ref{fs}) is not a free scalar field theory, since this
theory is periodic in the angle variable $\varphi$ (modulo $2\pi$).
Of course, if we neglect this periodicity and treat the variable
$\varphi$ as a non-compact variable 
$\varphi(x) \in (-\infty, +\infty)$, we have a trivial theory, i.e.,
free massless scalar field theory.  
In this case, the Contour integral is zero,
\begin{eqnarray}
 \oint_C \omega_\mu(z) dz^\mu =  {1 \over g} \oint_C  \partial_\mu
\varphi(z) dz^\mu = 0.
\end{eqnarray}
Hence, the Wilson loop  $W_c[\omega]$ is trivial,
$W_c[\omega] \equiv 1$ and hence the total static quark potential
comes from the the Wilson loop expectation  $W_c[v]$ of perturbative
U(1) gauge theory and is equal to the Coulomb potential.  Thus we
obtain a trivial result that the four-dimensional non-compact
Abelian gauge theory fails to confine quarks (charges). 
\par
However, the periodicity (or compactness) leads to  topological
non-trivial solutions which are seeds for confinement, as shown in
what follows.
\par
In the following, we restrict our consideration to $d:=D-2=2$ case.
The extremum of the classical action (\ref{fs}) is obtained as a
solution of the classical field equation,
\begin{eqnarray}
 \nabla^2 \varphi = 0 \quad ({\rm mod}~ 2\pi) .
\end{eqnarray}
The harmonic function $\varphi$ is constant or has singularities. 
We require $\varphi$ be constant at infinity and assume only isolated
singularities around which $\varphi$ varies by $\pm 2\pi$ as one
turns anticlockwise.  Then the solution
\footnote{
The two-dimensional Laplace equation is equivalent to the
Cauchy-Riemann equation.  Hence the solution is given by the
holomorphic function.  If one avoids branch cuts, it is a meromorphic
function. }
is written as
\begin{eqnarray}
 \varphi(z) = \sum_{i} Q_i \arctan {(z-z_i)_2 \over (z-z_i)_1}
 = \sum_{i} Q_i {\rm Im} \ln (z-z_i), \quad
 z := x_1 + i x_2 .
 \label{sol1}
\end{eqnarray}
This denotes a sum of vortex excitations located at points $x_i\in
R^2$ and of vorticity $Q_i$ (integers). 
The solution has an alternative form,
\begin{eqnarray}
 \varphi(z)  
 = \ln \prod_{i} {(z-z_i^+)/|z-z_i^+| \over (z-z_i^-)/|z-z_i^-|}
 = \sum_{i} \left[ \ln {(z-z_i^+) \over |z-z_i^+|}
 - \ln {(z-z_i^-) \over |z-z_i^-|} \right] .
\end{eqnarray}
This means that vortices of intensity $\pm 1$ are centered at the
points
$z_i^\pm$  and that intensities of higher magnitude are obtained when
several
$z_i^+$ (or $z_i^-$) coincide.
Note that the angle $\varphi$ is a multi-valued function, but
$e^{i\varphi}$ is well-defined everywhere, except at the singular
points. 
\par
The contribution of the solution (\ref{sol1}) to
$\omega_\mu$ is 
\begin{eqnarray}
 \omega_\mu(z) = {1 \over g} \partial_\mu \varphi(z) 
 = {1 \over g} \sum_{i} Q_i \epsilon_{\mu\nu} {(z-z_i)_\nu \over
(z-z_i)^2}  
 = {1 \over g} \sum_{i} Q_i \epsilon_{\mu\nu} \partial_\nu \ln
|z-z_i| .
\end{eqnarray}
Therefore, the integral of one-form $\omega := \omega_\mu dx^\mu$
along the closed loop $C$ is
\begin{eqnarray}
 \oint_C \omega_\mu dx^\mu = \oint_C \omega 
 = {1 \over g} \sum_{i} \int_{0}^{2\pi} d\Theta_i  
 =  \sum_{i} {2\pi \over g} Q_i,
\end{eqnarray}
where the sum runs over all the vortices inside the  
 closed loop  $C$ and $\Theta_i$ is an angle around $z=z_i$,
\begin{eqnarray}
 \Theta_i(z)  := \arctan {(z-z_i)_2 \over (z-z_i)_1} .
\end{eqnarray}
Note that $\omega$ is a closed form, $d\omega=0$, but it is not an
exact form, that is, there does not exist a function (zero-form)
such that $\omega=df$ with $f$ being defined everywhere in
$R^2-\{0\}$.  Domain of $f(z)=\Theta(z)$ is restricted to 
$R^2-R_+$, in other words, for one unit vortex at the origin,
\begin{eqnarray}
 \omega_\mu = \epsilon_{\mu\nu} {x_\nu \over x^2} 
 - 2\pi \theta(x_1) \delta(x_2) \delta_{\mu2}.
 \label{sing}
\end{eqnarray}
This is analogous to the case of the magnetic monopole in three
dimensions where the magnetic field is given by
\begin{eqnarray}
 H_\mu = {1 \over 2}{x_\mu \over |x|^3} - 2\pi \delta_{3\mu}
 \delta(x_1) \delta(x_2) \theta(x_3) .
\end{eqnarray}
The singular line of (\ref{sing}) in two dimensions does not
contribute to the action, so does the Dirac string (on the positive
$Z$ axis) in three dimensions.
\par
In order to calculate the classical action for the singular
configuration (\ref{sol1}), we consider a disk of radius
$R_0$, $D_i:=\{|z-z_i|<R_0; z\in R^2\}$ (centered on each singular
point $z_i$) which is small with respect to the distances between
vortices. Let
${\cal R}$ be the remaining domain of integration outside the
vortices.  The classical action consists of two parts,
the self-energy (action) part of vortices,
\begin{eqnarray}
 S^{(1)} = {\beta \over 2} \sum_{i} \int_{|z-z_i|<R_0} d^2z (\nabla
\varphi(z))^2 ,
\end{eqnarray}
and the remaining part,
\begin{eqnarray}
 S^{(2)} &:=& {\beta \over 2} \int_{{\cal R}} d^2z (\nabla
\varphi(z))^2
 \nonumber\\
 &=& - 2\pi \beta \sum_{i\not=j} Q_i Q_j \ln |z_i - z_j| 
 + \sum_{i} Q_i^2 \pi \beta \ln 1/R_0 .
\end{eqnarray}
Summing over all vortex sectors leads to the
partition function of the form,
\begin{eqnarray}
 Z_C 
 = \sum_{n=0}^\infty {\zeta^n \over (n!)^2}  \int \prod_{j=1}^n d^2
z_j
 \exp \left[ (2\pi)^2 \beta \sum_{i,j} Q_i Q_j \Delta(z_i,z_j)  
\right] ,
\quad \zeta := e^{-S^{(1)}} ,
\end{eqnarray}
where $\zeta$ comes from the self-energy (action) part of vortices  
and $\Delta$ expresses the two-dimensional inverse Laplacian given by
\begin{eqnarray}
  \Delta (x,0) = {1 \over 2\pi} \ln {R \over |x|} .
\end{eqnarray}
Therefore the partition function just agrees with the two-dimensional
neutral Coulomb gas (i.e., a gas of classical charged particles with
a Coulomb interaction and globally neutral, 
$\sum_{i} Q_i=0$).

\par
The transition temperature is estimated as follows.  The contribution
to the free energy from one vortex pair at distance
$r_{12}$ in a box of linear dimension $L$ is 
\begin{eqnarray}
 F &\sim& \ln \int_{|z_1 - z_2|>R_0} d^2z_1 d^2 z_2 \exp \left(  -
2\pi \beta  \ln |z_1 - z_2|/R_0 \right)
 \nonumber\\
 &\sim& \ln [ L^4 \exp \left(  - 2\pi \beta 
 \ln L/R_0 \right)]
 \nonumber\\
 &\sim& (4-2\pi \beta) \ln L .
\end{eqnarray}
The vortices always arise in pairs of opposite Coulomb charges to
yield finite energy configurations and each pair forms an elementary
dipole (Coulomb dipole gas).
If $2\pi \beta>4$ (the low temperature or weak coupling phase), the
contribution from the vortex pair is negligible in the limit 
$L \rightarrow \infty$.
In this phase, charges are bound and one has a
dielectric medium.  
As $\beta \rightarrow \infty$, few vortices are present and their
correlation decreases rapidly with the relative distance.  This
describes a dielectric medium of neutral bound states.
  In this regime, the correlation 
$\langle U(z)U(z') \rangle$ decays polynomially,
\begin{eqnarray}
 \langle U(z)U(z') \rangle =  |z-z'|^{-{1 \over 4\pi\beta}}.
\end{eqnarray} 
On the other hand, if $2\pi \beta<4$ (the high temperature or strong
coupling phase), an instability occurs  and the creation of
well-separated vortices is favored and disorder increases.
In high temperature phase, one has a plasma of free charges.
The vortex expectation decays exponentially
yielding exponential decay of 
$\langle U(z)U(z') \rangle$, 
\begin{eqnarray}
 \langle U(z)U(z') \rangle =  |z-z'|^{-{1 \over 4\pi\beta}}
e^{-m(\beta)|z-z'|}.
\end{eqnarray} 
Hence a naive estimate of the critical temperature is obtained
\begin{eqnarray}
 \beta_c = {2 \over \pi}, \quad g_c^2 = \pi^2 .
 \label{cp}
\end{eqnarray}
This is the phase transition without the appearance of a spontaneous
magnetization.  
The phase transition can be interpreted as a dipole condensation. 
The critical point separates the dissociated dipole phase from the
condensed phase.

\par
In order to calculate the Wilson loop, we use the equivalence of the
Coulomb gas to the sine-Gordon model (see Appendix A for a proof).  
The  sine-Gordon model is
defined by the action and the partition function,
\begin{eqnarray}
  S_{sG}(\phi) &:=& \alpha \int d^d x \left[ {1 \over 2}
(\partial_\mu \phi(x))^2
   - h \cos \phi(x) \right] ,
   \\
  Z_{sG}(h) 
     &:=& \int [d\phi] \exp [-S_{sG}(\phi) ] .
\end{eqnarray}
This is equivalent to the partition function of a globally neutral
gas of particles of charges $Q_i=\pm 1$ through a Coulomb potential
in $d$ dimensions,
\begin{eqnarray}
   Z_{C,\pm 1}(h) 
&=& \sum_{n=0}^\infty {z^{2n} \over (n!)^2}  
 \prod_{i=1}^n \int d^d x_i d^d y_j
\exp \Big\{ -{1 \over T} \Big[ \sum_{i<j} [ V(|x_i-x_j|) +
V(|y_i-y_j|)]
\nonumber\\
&& \quad \quad \quad \quad \quad -  \sum_{i,j}
V(|x_i-y_j|) \Big] \Big\} 
\end{eqnarray}
at temperature $T$ with the fugacity $z$,
\begin{eqnarray}
   {\alpha} = T = {1 \over 4\pi^2 \beta} = {g^2 \over 8\pi^3} , 
   \quad  
\alpha h = 2z = 2\zeta .
\end{eqnarray}
Note that $\alpha \beta = 1/(4\pi^2)$.
It is known that the transition point of sine-Gordon model is
\begin{eqnarray}
   {\alpha}_c =  {1 \over 8\pi}  , 
\end{eqnarray}
which is in agreement with (\ref{cp}). 
The relation of $\varphi$ and $\phi$ is given by
\begin{eqnarray}
   \phi(x,t) = \int_{x}^\infty dy \dot \varphi (y,t) ,
\end{eqnarray}
or
\begin{eqnarray}
   \partial_\mu \varphi = \epsilon_{\mu\nu} \partial^\nu \phi .
\end{eqnarray}
This implies that the fields $\varphi$ and $\phi$ are dual
variables.  Therefore, O(2) NLSM is equivalent to the Coulomb gas
and moreover it is equivalent to the sine-Gordon model when charge of
the Coulomb gas (or vorticity of O(2) NLSM) is restricted to
$Q_i=\pm1$.  Taking into account
$Q_i>1$ will lead to the  $\cos (Q \phi)$ term.
The above consideration can be transferred into the lattice
formulation, see \cite{Kogut79}.
\par
\subsection{Wilson loop and area decay}

\par
The Wilson loop expectation is calculated  using the
equivalent sine-Gordon model.
The generating functional for the charge density $\rho$ 
\begin{eqnarray}
  \rho(x) &:=& \sum_{i} Q_i \delta^{(2)}(x-x_i) ,
\end{eqnarray}
is obtained
as
\begin{eqnarray}
 {Z_{sG}[\eta] \over Z_{sG}(0)} &=& 
 \langle e^{i \int d^2x \rho(x) \eta(x)} \rangle_{sG} =
 \langle e^{i \sum_{i} Q_i \eta(x_i)} \rangle_{sG} ,
\\
 Z_{sG}[\eta]
  &=&  \int [d\phi] \exp \left\{ -\alpha \int d^2 x \left[ {1 \over
2} (\partial_\mu \phi(x))^2
   - h \cos [\phi(x)+\eta(x)] \right] \right\} .
   \label{Wle}
\end{eqnarray}
\par
In two dimensions, we can introduce the dual vector field,
\begin{eqnarray}
 H_\mu = \epsilon_{\mu\nu} \omega_\nu .
\end{eqnarray}
Then the dual field is connected with the charge density $\rho$ as
follows,
\begin{eqnarray}
  \oint_C \omega_\mu(z) dz^\mu = \int_{S} \epsilon_{\mu\nu}
\partial_\mu \omega_\nu(z) d^2z = \int \partial_\mu H_\mu(z) d^2z 
= {2\pi \over g} \int \rho(z) d^2z.
\end{eqnarray}
Note that the rotation of $\omega_\mu$ or the divergence of $H_\mu$ 
\begin{eqnarray}
  \epsilon_{\mu\nu} \partial_\mu \omega_\nu
= \partial_\mu H_\mu ={2\pi \over g} \rho
\end{eqnarray}
measures the density of topological charge. 
If we identify the right-hand-side with the magnetic charge, this
implies Dirac quantization condition,
\begin{eqnarray}
  g_m = {2\pi \over g} Q \quad (Q: {\rm integer}).
\end{eqnarray}
The Wilson loop is calculated from the $\eta$ given by
\begin{eqnarray}
  \eta(x) =  {q \over g} \oint_C dz^\mu \epsilon_{\mu\nu} 
  {(z-x)_\nu \over (z-x)^2} ,
\end{eqnarray}
since
\begin{eqnarray}
  q \oint_C \omega_\mu dz^\mu = \int d^dx \rho(x) \eta(x)
  = \sum_{i} Q_i \eta(x_i) .
\end{eqnarray}
Note that $\eta(x)=0$ if the argument $x$ of $\eta(x)$ is outside the
loop,   while $\eta(x)=2\pi q/g$ if $x$ is inside the loop.
\par
In high temperature phase, photon is massive, whereas photon is
massless in the low temperature phase.  This is because in the high
temperature phase, random distribution of free vortices with long
range interaction spoil the correlation. Therefore, in the high
temperature phase, the Wilson loop  expectation is estimated by the
steepest descent as
\begin{eqnarray}
  \langle W_C[\omega] \rangle_{sG} \cong
   \exp \left\{ -\alpha \int d^2 x \left[ {1 \over
2} (\partial_\mu [\phi_{cl}(x)-\eta(x)])^2
   - h \cos \phi_{cl}(x) \right] \right\},
\end{eqnarray}
where $\phi_{cl}$ is determined by the Debye equation,
\begin{eqnarray}
  \nabla^2 [\phi_{cl}(x)-\eta(x)]  =  h \sin \phi_{cl}(x) .
\end{eqnarray}
Corrections to this field due to fluctuation are exponentially small
in high temperature phase.
The loop $C$ is placed in two-dimensional plane.
We can perform the calculation in the same way as done by Polyakov
\cite{Polyakov77}.
\par
Instead of repeating similar calculation to
\cite{Polyakov77}, we use the Villain form \cite{Villain75},
\begin{eqnarray}
 e^{J \cos \phi} \rightarrow e^J \sum_{m \in Z} e^{-{J \over 2}
 (\phi-2\pi m)^2} ,
\end{eqnarray}
to estimate the Wilson loop expectation.  Then the partition
function is replaced with (apart from field-independent constants)
\begin{eqnarray}
 Z_{sG}[\eta]
&=&  
\int [d\phi] e^{-\alpha \int d^d x \left[ {1 \over
2} (\partial_\mu \phi(x))^2 \right] }
   \prod_{x\in R^d} e^{  \alpha h \cos [\phi(x)+\eta(x)] }
\nonumber\\   &=&  
\int [d\phi] e^{-\alpha \int d^d x \left[ {1 \over
2} (\partial_\mu \phi(x))^2 \right] }
  \prod_{x\in R^d} \sum_{m(x) \in Z} e^{-{\alpha h \over 2} 
[\phi(x)+\eta(x)-2\pi m(x)]^2}  
\nonumber\\   &=&  \sum_{\{m(x) \in Z ;x\in R^d\}}
\int  [d\phi] e^{-\alpha \int d^d x  {1 \over
2}  [- \phi(x) \partial^2 \phi(x)]  }
   e^{-{\alpha h \over 2} \int d^d x
[\phi(x)+\eta(x)-2\pi m(x)]^2}  
\nonumber\\   &=&  
\sum_{\{m(x) \in Z ;x\in R^d\}}  e^{-{\alpha h \over 2}
\int d^d x [\eta(x)-2\pi m(x)]^2}
\nonumber\\
&& \times
\int  [d\phi] e^{
 -  \int d^d x \{{\alpha \over 2} \phi(x) (-\partial^2+h)
\phi(x)    
  + \alpha h \phi(x)[\eta(x)-2\pi m(x)] \} }  
\nonumber\\   &=&  
\sum_{\{m(x) \in Z ;x\in R^d\}}  
\exp \Big[  - {\alpha h \over 2} \int d^d x \{ 
[\eta(x)-2\pi m(x)]^2 
\nonumber\\&&  
  - h [\eta(x)-2\pi m(x)]
  (-\partial^2+h)^{-1}[\eta(x)-2\pi m(x)] \} \Big]  .
\label{Wle2}
\end{eqnarray}
When $\eta=0$, the denominator is obtained,
\begin{eqnarray}
 Z_{sG}[0]
=
\sum_{\{m(x) \in Z ;x\in R^d\}}  
 \exp \left[  -(2\pi)^2 {\alpha h \over 2} \int d^d x \{   m(x)^2 
  -  h m(x)(-\partial^2+h)^{-1}  m(x)   \} \right] . 
 \label{Wle22}
\end{eqnarray}
Note that the field $\eta(x)$ has its support on $S(\partial S=C)$
and has the value $2\pi(q/g)$.
If $q$ is an integral multiple of $g$ (the elementary charge), we
have
$\eta \in 2\pi Z$.  This is absorbed by the shift of $m$. 
Therefore, in this case, charge confinement does not occur.  This is
interpreted as the charge screening.
\par
A naive estimate of the ratio
$Z_{sG}[\eta]/Z_{sG}[0]$
is given when $\eta \notin 2\pi Z$ in Appendix B.
Finally we obtain the area decay of the Wilson loop expectation,
\begin{eqnarray}
  \langle W_C[\omega] \rangle_{sG} &\cong& e^{-\sigma A(C)} ,
  \label{WlsG}
  \\
  \sigma &=& \left(2\pi{q \over g}\right)^2 {\alpha h \over 2} 
  = \left(2\pi{q \over g}\right)^2 \zeta \sim e^{-S^{(1)}} .
\end{eqnarray}
This implies the linear static potential 
\begin{eqnarray}
   V(R) = \sigma R  
   \label{lp}
\end{eqnarray} 
between two fixed electric
charges and electric string with uniform energy density
$\sigma$ which is called the string tension.
Therefore condensation of topological non-trivial configuration
leads to quark confinement.
From (\ref{iden}), (\ref{Cp}) and (\ref{lp}), the total static
potential is given by
\begin{eqnarray}
   V(R) = \sigma R  - {g^2 \over 4\pi}{1 \over R} + {\rm constant} . 
\end{eqnarray} 

\par
Even the continuum Abelian U(1) gauge theory has a confinement phase,
exhibiting rich phase structure as in lattice compact U(1) gauge
theory. 

\par

\section{Discussion}

In the four-dimensional pure U(1) gauge theory, we have proved the
existence of a strong coupling phase where the fractional
electric charge is confined by the linear static potential due to
vortex condensation.  In the following we discuss a few points of
perspective.

\subsection{Renormalization group and non-Gaussian fixed point}

\begin{figure}
\begin{center}
\unitlength=1cm
 \begin{picture}(12,6)
  \put(0,5.5){\bf $\beta(g)$}

  \put(4,2.4){$g_c$}
  \put(0,-0.5){\vector(0,1){5.5}}
  \put(0,2){\vector(1,0){9}}
  \put(-0.5,1.9){0}
  \put(9.5,1.9){$g$}
\thicklines
  \put(0,1.97){\line(1,0){4}}
\bezier{200}(4,1.97)(8.5,1.9)(9.5,-0.5)
 \end{picture}
\end{center}
 \caption{Renormalization group beta function of U(1) gauge theory}
 \label{fig1}
\end{figure}
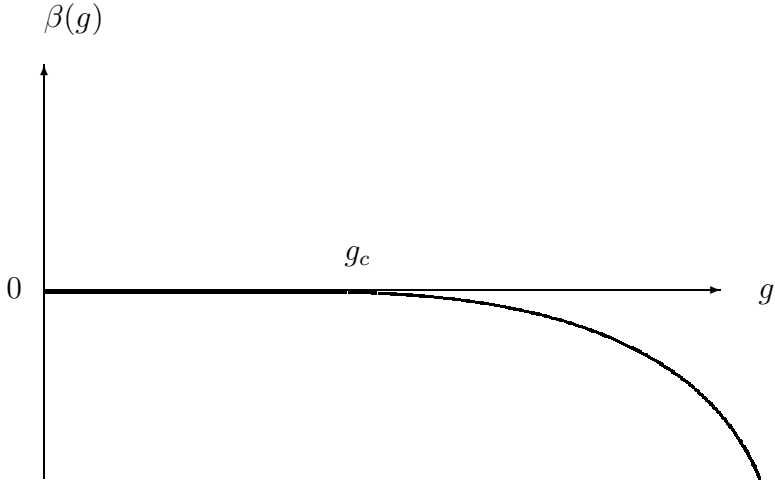
\par

The O(N) NLSM has the renormalization group beta function,
\begin{eqnarray}
  \beta(g) := \mu {dg(\mu) \over d\mu} 
  = - {N-2 \over 8\pi^2}g^3 + O(g^5)  .
\end{eqnarray}
In the low temperature ($0<T<T_c$), therefore, the O(2) NLSM has
vanishing beta function,
$
 \beta  \equiv 0 \quad (T \ll 1),
$
and $0<T<T_c$ is the line of fixed points.
This is consistent with the fact that at low temperature, the inverse
correlation length or mass $m=\xi^{-1}$ of O(2) NLSM or XY model
vanishes, i.e., $m(T) \equiv 0$ for all
$T<T_c$.  
The theory is conformal invariant.
For high temperature ($T>T_c$), the
renormalization group study of XY model shows that the mass behaves
as
\begin{eqnarray}
  m(T) \sim \exp \left(-{C \over \sqrt{T-T_c}} \right)  , \quad
  T \downarrow T_c ,  
  \label{BKTmass}
\end{eqnarray}
with a constant $C$.
\par
These results established in two-dimensional model would be
translated into the four-dimensional Abelian gauge theory, provided
that two theories have the same renormalization-group beta function.
We assume that in two theories  the mass
$m(g)$ is generated by the dimensional transmutation in such a way
that the beta function $\beta(g)$ is
related to the mass $m(g)$ through the well-known relation,
\begin{eqnarray}
  m(g) = \mu f(g) 
  =  \mu  \exp \left( - \int^g {dg \over \beta(g)} \right) ,
\end{eqnarray}
where $\beta(g)$ is defined by
\begin{eqnarray}
  \beta(g) := \mu {dg(\mu) \over d\mu} = - {f(g) \over f'(g)} .
  \label{betaf}
\end{eqnarray}
In the weak coupling phase ($g<g_c$), 
the gauge field is massless, $m(g) \equiv 0$ and the beta function of
the renormalization group \cite{WK74} is identically zero, 
\begin{eqnarray}
  \beta(g) \equiv 0 \quad (0<g<g_c) .
\end{eqnarray}
Therefore, $0<g<g_c $
is the line of fixed points. 
In strong coupling phase ($g>g_c$), the beta function  
behaves as (see Figure~1)
\begin{eqnarray}
  \beta(g)  = - {1 \over Bg_c}
  (g^2-g_c^2)^{3/2} < 0 ,  \quad (g>g_c, g \cong g_c) ,
\end{eqnarray}
which is compatible with the BKT mass (\ref{BKTmass}), 
\begin{eqnarray}
  m(g) = A \exp \left(-{B \over \sqrt{g^2-g_c^2}} \right)  , \quad
  g \downarrow g_c ,  
  \label{BKTmass2}
\end{eqnarray}
with constants $A, B >0$.  Note that $\beta(g)$ is independent of
$A$.
\par
It is worth remarking that the recent lattice computer simulation
\cite{JLN96,CFJLNS97} indicates the existence of an non-Gaussian
fixed point in four-dimensional pure compact U(1) gauge theory.
This should be compared with the old results \cite{EJNZ85,JNZ85}.
In the simulation \cite{JLN96,CFJLNS97} the continuous phase
transition was found and  analyzed  according to the power law
scaling, although our investigation suggests the scaling behavior of
the essential singularity type and the data do not exclude the
essential singularity.  It will be rather difficult to
specify the essential singularity in the computer simulations, since
the lattice size available is not yet so large to confirm this
issue.  Anyway, it will be interesting to find any relationship to
fill the gap between two approaches.  
\par
If the mass scale is generated by dimensional transmutation,
the above results are quite analogous to the situation found for the
dynamical fermion mass and the beta function in quenched
massless QED
\cite{Miransky85,KMY89}. The relationship of these results with the
quenched QED is more suggestive by the method of bosonization or
fermionization.
The two-dimensional sine-Gordon model is equivalent to the massive
Thirring model \cite{Coleman75,Kondo95} with an action,
\begin{eqnarray}
  S[\psi, \bar \psi] 
  = \int d^2x [\bar \psi(i\gamma_\mu \partial_\mu + m) \psi
  - {G \over 2}(\bar \psi \gamma_\mu \psi)^2 ] .
\end{eqnarray}
The correspondence between two theories is given by
\begin{eqnarray}
 1 + {G \over \pi} &=& 4\pi \alpha,
 \\
  \bar \psi \gamma_\mu \psi 
  &=& - {1 \over 2\pi} \epsilon_{\mu\nu} \partial_\nu \phi,
  \\
   m \bar \psi \psi &\rightarrow& - \alpha h \cos \phi .
\end{eqnarray}
Our result shows that the transition point $\alpha_c = 1/(8\pi)$
corresponds to
$G_c=-\pi/2$. 
By making use of this equivalence, the Wilson loop (\ref{WlsG}) in
the topological non-trivial sector of the four-dimensional U(1)
gauge theory can be calculated in the two-dimensional massive
Thirring model. The details will be given elsewhere.

\subsection{Lower and higher dimensional cases}

Using the equivalence between the TQFT part of U(1)$_D$ gauge theory
and O(2)$_{D-2}$ NLSM, we can study other dimensional cases.
\par
For $D=3$, the equivalent O(2) NLSM is one dimensional, O(2)$_1$. 
This is not a field theory model, but a quantum mechanical model of
the plane rotor.  There is no phase transition in this model. 
This implies that the three-dimensional U(1) gauge theory has
only the confinement phase.  This will be understood as the
tunneling effect among classical vacua in the sense that the
double-well anharmonic oscillator is related to the
one-dimensional Ising model.   It will be interesting to see
agreement (or disagreement) on confinement mechanism between our
approach and the Polyakov approach
\cite{Polyakov75,Polyakov77}.

\par
For $D=5$, the O(2) NLSM is three dimensional, O(2)$_3$.  The
three-dimensional O(2) NLSM has two phases on the
lattice \cite{TS78}.  The phase transition is first order.
This result is consistent with the mean field study of
five-dimensional lattice U(1) gauge theory \cite{DZ83}.
This theory has a finite non-zero critical coupling $g_c$. 
In the strong coupling phase, quark confinement is expected
to occur.  However, the phase transition is first order. 
Therefore, on the lattice, it is impossible to take the continuum
limit at this point.  In view of this, the construction of continuum
U(1) gauge theory from the lattice regularized theory is problematic
in five dimensions and higher dimensions, unless the action is
modified.

\section*{Acknowledgment}
The author would like to thank Jiri Jersak
and Volodya Miransky for informing him of
the recent results of the computer simulation 
\cite{JLN96,CFJLNS97}.
This work is supported in part by the Grant-in-Aid for
Scientific Research from the Ministry of Education, Science and
Culture.

\appendix
\section{Equivalence between Coulomb gas and sine-Gordon model}
\setcounter{equation}{0}

The partition function  can be rewritten as
\begin{eqnarray}
 Z_{sG}(h) &=& \int [d\phi] e^{-\alpha \int d^d x  {1 \over 2}
(\partial_\mu \phi(x))^2} 
 \sum_{n=0}^\infty {(\alpha h)^n \over n!} \left[ \int d^d x \cos
\phi(x)\right]^n
\nonumber\\
 &=& \int [d\phi] e^{-\alpha \int d^d x  {1 \over 2}
(\partial_\mu \phi(x))^2} 
 \sum_{n=0}^\infty {(\alpha h)^{2n} \over (2n)!} {1 \over 2^{2n}}
\left[ \int d^d x 
 (e^{i\phi(x)}+e^{-i\phi(x)}) \right]^{2n}
\nonumber\\
 &=& \int [d\phi] e^{-\alpha \int d^d x  {1 \over 2}
(\partial_\mu \phi(x))^2} 
 \sum_{n=0}^\infty {(\alpha h)^{2n} \over (2n)!}  
 {1 \over 2^{2n}} \pmatrix{2n \cr n} \prod_{i=1}^n 
 \int d^d x_i d^d y_i e^{i\phi(x_i)-i\phi(y_i)} 
\nonumber\\
 &=& \int [d\phi] e^{-\alpha \int d^d x  {1 \over 2}
(\partial_\mu \phi(x))^2} 
 \sum_{n=0}^\infty {(\alpha h/2)^{2n} \over (n!)^2}  
 \prod_{i=1}^n 
 \int d^d x_i d^d y_i e^{i\phi(x_i)-i\phi(y_i)} .
\end{eqnarray}
Note that
\begin{eqnarray}
   [[ e^{\int d^dx J(x) \phi(x)} ]] &:=& \int [d\phi] 
   \exp \left[ - {\alpha \over 2} \int d^d x (\partial_\mu
\phi(x))^2 + \int d^dx J(x) \phi(x) \right] 
\nonumber\\
&=& \exp \left[  {1 \over 2\alpha} \int d^d x \int d^dy 
J(x) \Delta(x,y) J(y) \right]  ,
\end{eqnarray}
where $\Delta(x,y)$ is the massless scalar field propagator,
\begin{eqnarray}
 \Delta(x,y) := \int {d^dp \over (2\pi)^d} {e^{ip(x-y)} \over p^2} .
\end{eqnarray}
In particular, for 
$
 J(x) = i \sum_{j} q_j \delta(x-x_j) ,
$
\begin{eqnarray}
   [[ \prod_{j=1}^n e^{i q_j \phi(x_j) } ]] = 
   \cases{ \exp \left[ - {1 \over 2\alpha}
\sum_{j,k} q_j q_k \Delta(x_j,x_k)  \right]  
&  for  $\sum_{j} q_j = 0$ \cr
0 & for $\sum_{j} q_j \not= 0$} ,
\label{formulab}
\end{eqnarray}
where the latter case is a result of invariance under constant
translation of the field $\varphi(z) \rightarrow \varphi(x) + c$.
Using this result for $q_j=\pm1$, 
\begin{eqnarray}
   Z_{sG}(h) 
&=&  \sum_{n=0}^\infty {(\alpha h/2)^{2n} \over (n!)^2}  
 \prod_{i=1}^n \int d^d x_i d^d y_i
 \int [d\phi] e^{-\alpha \int d^d x  {1 \over 2}
(\partial_\mu \phi(x))^2}  \prod_{i=1}^n 
 e^{i\phi(x_i)-i\phi(y_i)} 
\nonumber\\
 &=& 
 \sum_{n=0}^\infty {(\alpha h/2)^{2n} \over (n!)^2}  
 \prod_{i=1}^n \int d^d x_i d^d y_i
\exp \Big\{ -{1 \over 2\alpha} \Big[ \sum_{i<j} [ \Delta(x_i-x_j) +
\Delta(y_i-y_j)]
\nonumber\\
&& \quad \quad \quad \quad \quad -  \sum_{i,j}
\Delta(x_i-y_j) \Big] \Big\} .
\end{eqnarray}

\section{Estimate of sine-Gordon partition function}
\setcounter{equation}{0}

\par
Note that the quantity
\begin{eqnarray}
  \varrho(x)^2  -  h \varrho(x)(-\partial^2+h)^{-1}\varrho(x) 
  = \varrho(x) {-\partial^2 \over -\partial^2+h} \varrho(x)
\end{eqnarray}
is positive, since 
$(-\partial^2)/(-\partial^2+h)$ is a positive operator.
Therefore, 
\begin{eqnarray}
   e^{-{\alpha h \over 2}(2\pi)^2  \int_S d^d x \{ \varrho(x)^2 
  - h \varrho(x) (-\partial^2+h)^{-1} \varrho(x)  \} },
  \quad
  \varrho(x) = \Big|{\eta(x) \over 2\pi}-m(x) \Big|
\end{eqnarray}
is monotonically (rapidly) decreasing in $\{ |\rho(x)|; x \in S \}$.
Therefore, in the partition function,
\begin{eqnarray}
 Z_{sG}[\eta]
=   \sum_{\{m(x) \in Z ; x\in R^d\}}  
\exp \Big[  - {\alpha h \over 2} (2\pi)^2 \int d^d x \{ \varrho(x)^2 
  - h \varrho(x)
  (-\partial^2+h)^{-1}\varrho(x) \} \Big] ,
\end{eqnarray}
the most dominant contribution comes from a set of configurations 
$\{ |\varrho(x)|; x \in S \}$
which gives the smallest value for
$
\int d^d x \{ \varrho(x)^2 - h \varrho(x)
  (-\partial^2+h)^{-1}\varrho(x) \} .
$ 

\par
If the argument $x$ of $\eta(x)$ is outside the loop
$C( C=\partial S)$, $\eta(x)=0$, while $\eta(x)=2\pi
q/g$ if $x$ is inside the loop. For $Z_{sG}[0]$,  $\{ m(x) \equiv 0
\}$ gives the most dominant contribution.  Hence, we see
\begin{eqnarray}
Z_{sG}[0] = 1 + \cdots .
\end{eqnarray}
For $Z_{sG}[\eta]$, the most dominant contribution comes from a set
of integers $\{ m(x) \}$ whose value is the nearest to $q/g$
where $\varrho(x)=2\pi|q/g-m(x)|$.  
Since the integral part of $q/g$ is absorbed in the shift of $m(x)$, 
it is sufficient to consider the case $0<q/g<1$ without loss of
generality.
For a half-integer $q/g$,  i.e., $q/g=\pm 1/2$, we see that
$\{ m(x) \equiv 0, \pm 1 \}$ give the smallest value of 
$2\pi|q/g-m(x)|$ for $x \in S$.
For $0<q/g<1/2$ (resp. $1/2<q/g<1$), the most dominant contribution
is given by  $\{ m(x) \equiv 0 \}$ (resp. $\{ m(x) \equiv 1 \}$).
Thus, the rough estimate, for example, in the case of $0<q/g<1/2$ 
leads to 
\begin{eqnarray}
 Z_{sG}[\eta]
  &=&  
  e^{-{\alpha h \over 2}  \int_S d^d x \{ \eta(x)^2 
  - h \eta(x) 
  (-\partial^2+h)^{-1} \eta(x)  \} }
  + \cdots  .
\end{eqnarray}
\par
A naive estimate of the ratio
$Z_{sG}[\eta]/Z_{sG}[0]$
is given for $\eta \notin 2\pi Z$ by
\begin{eqnarray}
 \langle W_C[\omega] \rangle_{sG} = {Z_{sG}[\eta] \over Z_{sG}[0]}
  &=&  
  e^{- {\alpha h \over 2} \int_S d^d x  \eta(x)[ 1
  - h (-\partial^2+h)^{-1} ]\eta(x)    }
  + \cdots  .
\end{eqnarray}
Here $(-\partial^2+h)^{-1}(x,y)$ is the massive scalar propagator
with mass $m \sim \sqrt{h}$ and hence has an exponential damping
factor 
$e^{-m|x-y|}$.  Therefore, the integral 
$\int_S d^d x \eta(x)(-\partial^2+h)^{-1} \eta(x)$
converges to a finite value even for large $S$.
The term $\int_S d^2 x  \eta(x)^2$ is proportional to the area of
$S$. This term leads to the area decay of the Wilson loop
expectation.

\end{document}